# Design and resonator-assisted characterization of high performance lithium niobate waveguide crossings

Yikun Chen,[1] Ke Zhang,[1] Hanke Feng,[1] Wenzhao Sun,[2,3] Cheng Wang[1,4]*

[1]Department of Electrical Engineering, City University of Hong Kong, Kowloon, Hong Kong, China

[2]Centre of Internet of Things, City University of Hong Kong Dongguan Research Institute, Dongguan, China

[3]Centre of Information and Communication Technology, City University of Hong Kong Shenzhen Research Institute, Shenzhen, China

[4]State Key Laboratory of Terahertz and Millimeter Waves, City University of Hong Kong, Kowloon, Hong Kong, China

*Corresponding author: cwang257@cityu.edu.hk

## Abstract

Waveguide crossings are elementary passive components for signal routing in photonic integrated circuits. Here, we design and characterize two multimode interferometer (MMI)-based waveguide crossings to serve the various routing directions in the anisotropic x-cut thin-film lithium niobate (TFLN) platform. To address the large measurement uncertainties in traditional cut-back characterization methods, we propose and demonstrate a resonator-assisted approach that dramatically reduces the uncertainty of insertion loss measurement(< 0.021 dB) and the lower bound of crosstalk measurement (-60 dB) using only two devices. Based on this approach, we demonstrate and verify TFLN waveguide crossings with insertion losses of < 0.070 dB and crosstalk of < -50 dB along all three routing directions at 1550 nm. The low-loss and low-crosstalk waveguide crossings in this work, together with the simple and efficient characterization strategy, could provide important layout design flexibility for future large-scale classical and quantum TFLN photonic circuits.

## Introduction

Photonics integrated circuits (PICs) are promising for next-generation communication and computing systems [1]. However, unlike electronic very large-scale integration (VLSI), which utilizes complex three-dimensional routing structures, e.g., vertical vias and horizontal interconnections, to realize low-loss and compact multilayer signal routing, PICs find difficulties in realizing such an optical counterpart, since optical via will suffer from high loss due to the sharp refractive index change at the vertical bends. Although multilayer waveguides could be implemented and linked via adiabatic tapers [2], the required fabrication processes are complicated and cost-ineffective. On the contrary, photonic interconnects could also cross over each other in a planar fashion, via two-dimensional waveguide crossings, which are intriguing since they need only a single lithography step and are easy to implement. They have been applied to realize compact optical modulators [3], optical computing systems [4], quantum systems [5], etc. The insertion losses and crosstalk of these waveguide crossings, although seemingly small on individual bases, are critical performance metrics, particularly for large-scale PICs, where vast amounts of waveguide crossings are adopted and their losses and crosstalk values cascade [4, 5]. While some people make efforts to reduce the number of waveguide crossings by optimizing the routing algorithm [6], it is still a key research theme to improve the performance of waveguide crossings [7-12].

To date, high-performance waveguide crossings have been achieved based on different principles, including inverse design [7], Gaussian beam synthesis [8], sub-wavelength grating [9], and multimode interference (MMI) [10-12], etc. MMI-based designs are the most widely adopted approach due to the ease of design, high fabrication robustness, and good performance. They have been shown on silicon [10], polymer [11], and silicon nitride [12] platforms. However, investigations on thin-film lithium niobate (TFLN) remain scarce, which is an emerging candidate for large-scale PICs, owing to its low optical loss, large nonlinear coefficient, and commercial large-size wafer availability (up to 6 inches). Benefiting from those properties, high quality factor ($Q$ factor) micro-resonators [13], high-speed electro-optic modulators [14], and broadband frequency combs [15] have been realized. The development of high-performance TFLN waveguide crossings is indispensable for routing and intermediating between these devices in future large-scale PICs. On the other hand, traditional cut-back characterization of waveguide crossings often sees substantial measurement uncertainties especially for small insertion losses, since the fiber-chip coupling losses (~ 5 dB/facet) [16] are usually much larger than the on-chip loss of an individual crossing (< 0.1 dB) and they vary substantially from device to device (commonly ± 0.4 dB). The difficulty in accurately determining the total optical loss is further exacerbated by the Fabry–Pérot interference patterns resulting from reflections at the chip facets in edge-coupling setups. As a result, these measurements typically require a large number of cascaded crossings and averaging over many devices for a reasonably reliable estimation. Accurate measurements of small crosstalk values of these crossings are also highly non-trivial since the measurement lower bound (noise floor) could easily be affected by scattered light from other parts of the chip.

In this letter, we address these issues by proposing a resonator-assisted measurement method, where the waveguide crossing under test is embedded in a high-$Q$ resonator. As a proof-of-concept, we design and fabricate low-loss MMI-based waveguide crossings in three routing crystal directions on an x-cut TFLN chip and precisely characterize the devices using our proposed method. Our experimental results show dramatically improved insertion loss measurement accuracy and lowered crosstalk measurement floor compared with traditional cut-back approaches.

## Results

Figure 1(a) schematically illustrates the traditional cut-back method and our proposed resonator-assisted insertion loss and crosstalk characterization methods. The traditional cut-back method compares the transmission spectra of multiple waveguides with different numbers of cascaded crossings to obtain the insertion loss from linear fitting [7, 8, 10-12], where the inaccurate estimation of fiber-chip coupling and the Fabry–Pérot fringes resulted from the chip facets both contribute to the final measurement uncertainty. In contrast, our resonator-based method compares the $Q$ factors of a reference micro-ring resonator and a crossing-embedded resonator to extract the difference in round-trip losses, which is not affected by uncertainties in fiber-chip coupling. This method is also not sensitive to fabrication variations between devices if the intrinsic loss is much lower than crossing loss. The effect of Fabry-Perot fringes can also be mitigated since we focus only on spectrally narrow resonance dips, especially for high-$Q$ resonators. For crosstalk characterization, the traditional method that compares the transmission level of through and cross ports of a single crossing requires a large input power to reveal small crosstalk values and can easily be affected by scattered light from other parts of the chip, limiting the minimum crosstalk value we can measure. On the contrary, our resonator-enhanced approach could effectively bring the crosstalk signal up from the noise floor by magnifying the crosstalk power in the high-$Q$ resonator.

To experimentally verify our proposed resonator-assisted measurement approaches, we first design two types of low-loss waveguide crossing to serve three popular routing directions in an anisotropic x-cut lithium niobate (LN) film, as shown in Figure 1(b). Specifically, the left crossing device aligns along the y/z crystal directions allowing z (0°)- and y (90°)-propagating signals to cross over each other with low crosstalk, which requires an asymmetric design due to the different refractive indices along y- and z-crystal directions. The right crossing device, on the other hand, routes along the 45° direction between y and z and can be achieved using a symmetric design. We numerically simulate the MMI structure using full 3D finite-difference time-domain (FDTD) simulation (Ansys Lumerical). The anisotropic refractive index of LN is adopted to simulate the insertion loss and the crosstalk along different routing crystal directions. The top width of the routing waveguide, the rib waveguide thickness, the slanted sidewall angle of the rib waveguide, and the slab thickness are fixed to be 1 μm, 250 nm, 67° and 250 nm, respectively, following our typical TFLN device parameters. Geometric parameters, including the length of taper $L_T$, the length of MMI region $L_C$, and the top width of MMI region $W_M$, are swept to achieve optimal designs with low loss and low crosstalk for fundamental transverse-electric ($TE_0$) mode in each design [Figure 1(c)]. For asymmetric design, these parameters are separately optimized for the 0° and 90° arms.

To serve as an example, Figure 2 shows the simulated insertion losses for the symmetric crossing at 1550 nm as functions of various $L_C$, $L_T$, and $W_M$ values. Red stars denote the optimal insertion loss of 0.038 dB with geometric parameters of $W_M$ = 3.3 μm, $L_C$ = 10.5 μm, and $L_T$ = 7 μm. The optimized asymmetric crossing has $W_M$, $L_C$, and $L_T$ values of 3.4 μm, 12.5 μm, 7 μm for the 90° arm, and 3.2 μm, 10 μm, 6 μm for the 0° arm. Devices are subsequently fabricated on a commercial x-cut LN-on-insulator wafer (NANOLN). A $SiO_2$ layer is first deposited. The pattern is then transferred to the $SiO_2$ layer using an ASML UV Stepper lithography system (NFF, HKUST), followed by a reactive ion etching (RIE) process. Next, a second step of RIE transfers the pattern to the LN layer. Finally, chips are cleaved for end-fire coupling[15].

Next, we illustrate the limitations of traditional cut-back measurements by characterizing the insertion losses of the fabricated TFLN crossings. Figure 3(a) shows the normalized measured transmission spectra of the waveguides with different numbers of 45°-routing crossings, measured using a broadband tunable laser (Santec TSL-550), followed by a polarization controller to launch $TE_0$ mode and a lensed fiber for efficient coupling. The output light is collected by another lensed fiber and finally transmitted to a photodiode (PD). The measured transmittance descends with the increasing number of crossings, whose trends can be fitted by a linear regression to estimate the insertion loss of a single crossing. Fabry–Pérot interference fringes can be clearly observed in these spectra, which adds a measurement uncertainty (standard deviation) of $\sigma_s$ = 0.510 dB. Figure 3(b) shows the linear regression of this device with a fitted loss of 0.022 dB at 1550 nm and an uncertainty of 0.056 dB. The total measurement uncertainty is estimated as $\sigma_t = [(\sigma_s^2 + \sigma_c^2)/c_x]^{1/2}$, where $\sigma_c$ = 0.368 dB is the standard deviation (red error bars) resulting from coupling and fabrication variation between different devices, $\Sigma_x = \Sigma(x_i - \bar{x})^2 = 125$ is the summed deviation of the crossing number ($x_i$) of each cut-back waveguide from the mean number ($\bar{x}$). We calibrate $\sigma_c$ through measuring the transmittances of four reference waveguides (without crossings) on the same chip by $\sigma_c = (\Sigma(T_i - \bar{T})^2/4)^{1/2}$, where $T_i$ is the transmittance of each reference waveguide, $\bar{T}$ is the average transmittance. Notably, the estimated measurement uncertainty of 0.056 dB is substantially larger than the estimated loss (0.022 dB) itself even though 5 separate devices with up to 20 crossings are tested, showing significant limitations in measuring small losses. By adopting the same method, we estimate the insertion loss of the two designs along three routing crystal directions in Figure 3(c) [black arrows in the insets indicate the directions with respect to optical axis (red dashed line)], where fitted insertion losses and measurement uncertainties are respectively given by blue dashed lines and widths of shaded areas. The estimated insertion losses along 45°, 90°, and 0° at 1550 nm are respectively 0.022 ± 0.056 dB, 0.109 ± 0.055 dB, and 0.122 ± 0.064 dB, all showing large measurement uncertainties > 0.050 dB.

Similar challenges are also present for small crosstalk measurements using traditional methods. The bottom gray curve in Figure 3(a) shows the crosstalk spectra. The measured crosstalk values (~ -40 dB) are significantly higher than our simulated results (~ -60 dB), which may result from scattered noises from defects on chip and from facets. The measurement is also substantially limited by a relatively high noise floor of ~ -45 dB in our current measurement system, which could be estimated from the maximum output power of 13 dBm of the laser, the noise floor of ~ -50 dBm of our PD, the fiber-chip coupling loss of ~ 5 dB per facet [15], and a few dB of additional losses in other parts of the fiber link. Figure 3(d) summarizes the measured and simulated crosstalk along the three routing directions, all showing substantially higher (> 15 dB) measured crosstalk values than the simulated ones.

Finally, we characterize our designed waveguide crossings using our proposed resonator-assisted method, showing significantly improved insertion loss measurement accuracy and lowered crosstalk measurement floor. Figure 4(a) shows example normalized intensity spectra of a crossing-embedded resonator along 45° routing direction (bright blue line) and a reference resonator (grey line) fabricated side by side on the same TFLN chip. Lorentzian fitting [right panel of Figure 4(a)] reveals a substantially lowered loaded $Q$ factor of 226,000 for the crossing-embedded resonator compared with that of the reference resonator (~ 300,000) due to the excessive loss induced by the crossing structure. We can extract the round-trip amplitude transmission coefficient $a$ of the resonators from the loaded $Q$ factors using the equation[16]:

$$a = 1 - \frac{\pi n_g L}{2\lambda Q_L} - \sqrt{\frac{1}{T_t}} + \sqrt{(\frac{\pi n_g L}{2\lambda Q_L})^2 - \frac{\pi n_g L}{\lambda Q_L} + \frac{1}{T_t}} \,, \tag{1}$$

where $n_g$ and $Q_L$ are the group index and loaded quality factor at the target wavelength $\lambda$, $L$ is the circumference of the resonator. $T_t = I_t/I_i$ is the on-resonance relative transmittance, where $I_i$ and $I_t$ are respectively the off-resonance and on-resonance light intensity. We can then estimate the insertion loss $IL$ of the crossing by comparing the round-trip intensity transmittances of the crossing-embedded resonator ($a_c^2$) and the reference resonator ($a_r^2$) using the equation:

$$IL = 10\log_{10}\frac{a_c^2}{a_r^2}. \tag{2}$$

Figure 4(b) shows the Lorentzian fitted loaded $Q$ factors along the three crossing directions. The fluctuations of $Q$ factors mainly come from the deformed Lorentzian shape affected by the background resonance fringes. Despite these fluctuations, we show much smaller insertion-loss measurement uncertainties [shaded areas in Figure 4(c)] than the cut-back results in Figure 3(c). The measurement uncertainty is calculated as $\sigma_t = (\sigma_s^2 + \sigma_c^2)^{1/2}$, where $\sigma_s$ is the standard deviation of extracted insertion losses [blue dots in Figure 4(c)], in this case < 0.018 dB, and $\sigma_c$ is the standard deviation due to fabrication variation between different resonators. We calibrate $\sigma_c$ through statistical analysis of the round-trip intensity transmittance of several reference resonators on the same chip, which is only as small as ~0.009 dB (0.368 dB in cut-back method). Importantly, all above analyses are not influenced by fiber-chip coupling variations, leading to small insertion loss measurement uncertainty ~ 0.02 dB with only two devices needed. The measured insertion losses along the three directions around 1550 nm are respectively 0.051 ± 0.014 dB, 0.065 ± 0.013 dB, and 0.070 ± 0.021 dB, showing improved and lowered uncertainties (> 0.055 dB in cut-back results) and good consistency with simulation results.

For crosstalk measurement, we monitor the output optical signals from the cross port of the crossing-embedded resonator, which shows a resonance-enhanced crosstalk peak $I_{cr}$ [dark blue line in Figure 4(a)] at each resonant wavelength. We can estimate the crosstalk of the crossing by using the equation:

$$Crosstalk = 10\log_{10}\frac{T_c}{M}, \tag{3}$$

where $T_c = I_{cr}/I_i$ is the on resonance relative crosstalk transmittance, $M$ is the intensity amplification factor of the resonator:

$$M = \frac{1-t^2}{(1-a_c t)^2}. \tag{4}$$

$t$ is the amplitude transmission coefficient of the waveguide coupler, which can be estimated from the loaded $Q$ factor of the resonator by:

$$t = 2 - a_c - \frac{\pi n_g L}{\lambda Q_c}. \tag{5}$$

Figure 4(d) illustrates the extracted crosstalk values of the three routing crystal directions, i.e. $-50.1 \pm 5.1$ dB, $-58.0 \pm 5.1$ dB, and $-56.2 \pm 2.8$ dB, respectively, which are in much better agreement with our theoretical prediction. Thanks to the ~15 dB power amplification in our resonator, we bring down the lowest measured crosstalk values from ~ -45 dB in the traditional method to ~ -60 dB here, leading to much more accurate measurements of the real crosstalk values.

In conclusion, we have demonstrated two MMI-based waveguide crossings along three routing directions in TFLN, achieving insertion losses of < 0.070 dB and crosstalk of < -50 dB at 1550 nm for all devices. The ability to cross over waveguides with low losses and crosstalk could provide layout-design flexibility for future large-scale TFLN PICs. Our designed waveguide crossings are characterized by a resonator-assisted method, which shows an improved insertion loss measurement accuracy of < 0.021 dB and lowered crosstalk measurement floor by 15 dB, compared with traditional cut-back methods. Both the insertion loss uncertainties and the minimum measurable crosstalk can be further reduced by increasing the $Q$ factors of the resonator through improved fabrication. The proposed resonator-assisted method provides an accurate and efficient solution to measure ultra-low loss waveguide crossings, and could be transferred to other photonic platforms.

## Acknowledgements

This work is supported in part by the Research Grants Council, University Grants Committee (CityU 11204820, CityU 11212721, N_CityU113/20); Croucher Foundation (9509055).

## Competing interests

The authors declare no competing interests.

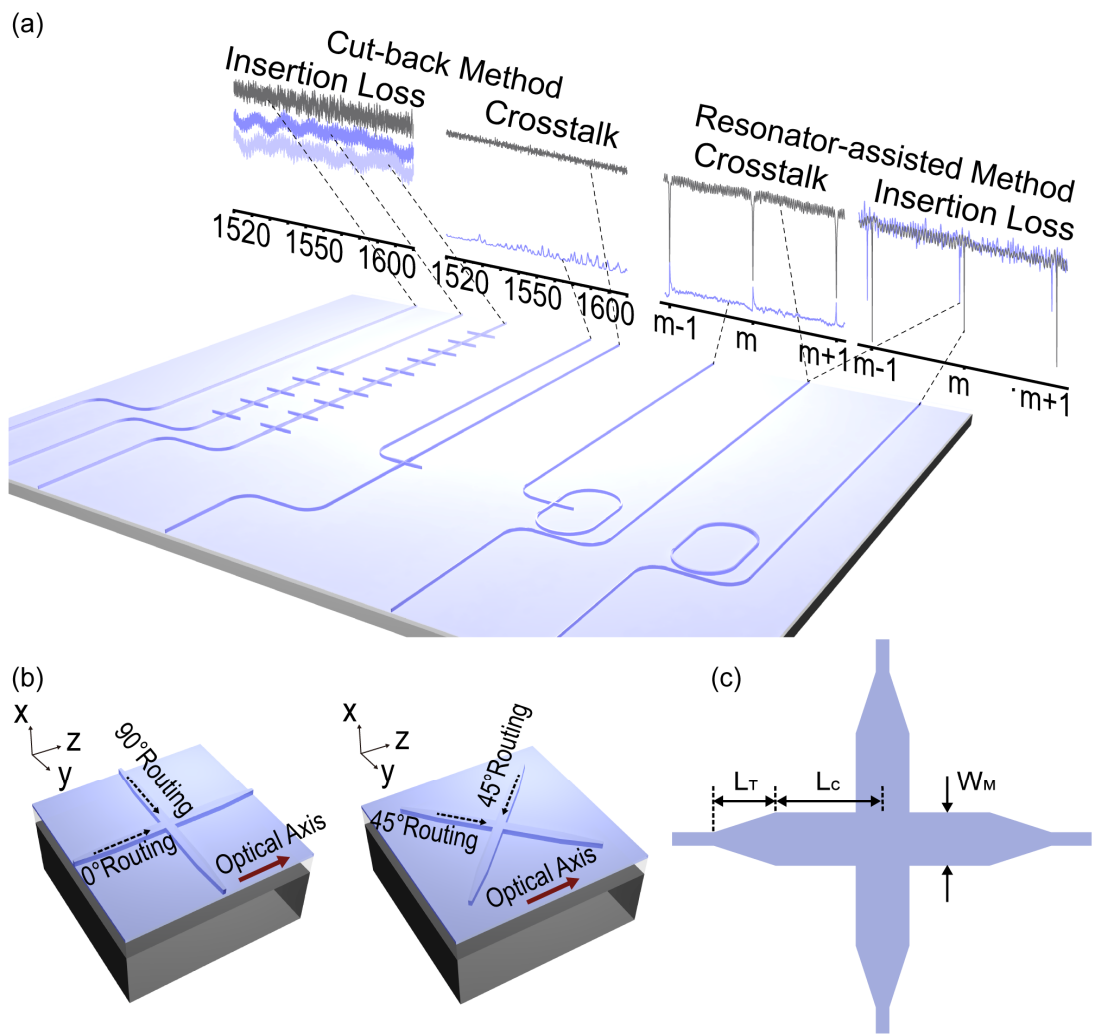

**Fig. 1.** (a) Illustration of cut-back and resonator-assisted measurements for insertion loss and crosstalk characterization. (b) Schematics of two designs for three crystal routing directions. (c) Geometric parameters swept for achieving optimal designs.

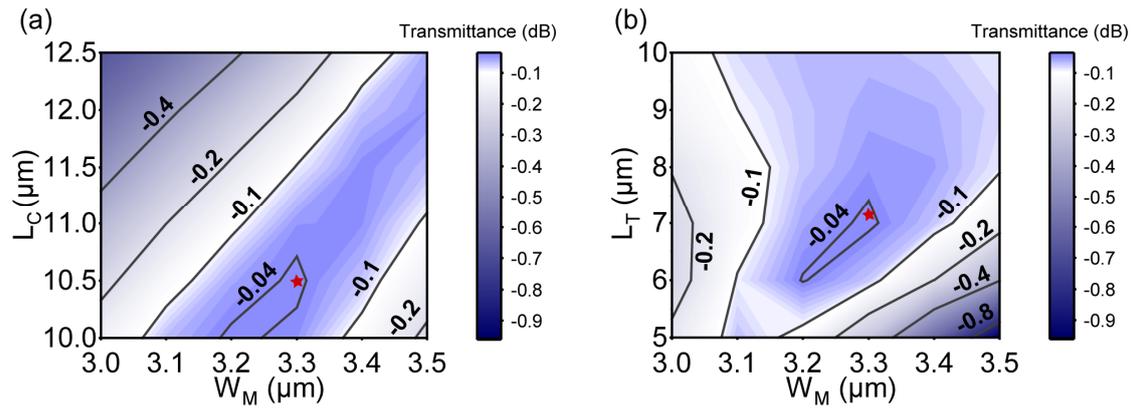

**Fig. 2.** (a-b) Simulated insertion loss as functions of geometric parameters of (a) $W_M$, $L_C$ and (b) $W_M$, $L_T$ along 45° routing crystal direction. Red stars denote optimal parameters used in our experiment.

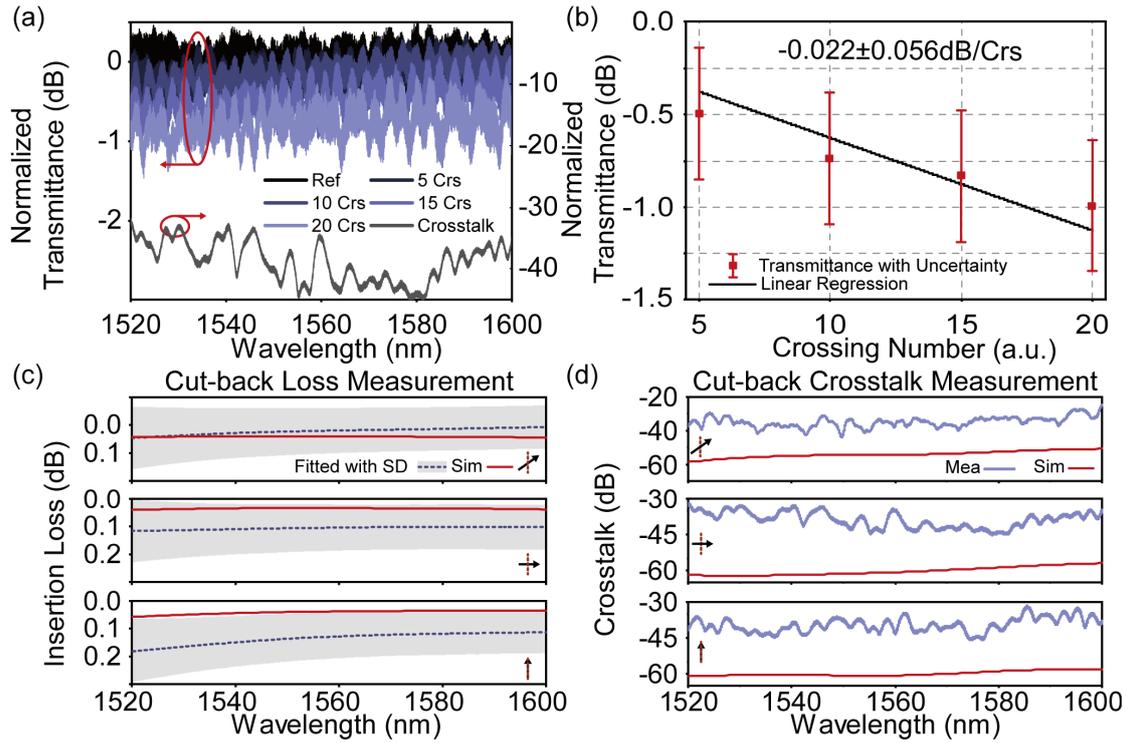

**Fig. 3.** Characterizations of TFLN crossings using traditional methods. (a) Measured transmission spectra for different numbers of cascaded 45° routing waveguide crossings (blue lines), and transmittance at the cross-port of a single waveguide crossing (bottom gray line). (b) Linear regression of insertion loss at 1550 nm. (c) Fitted insertion loss spectra (blue dashed lines) with standard deviation (SD, shaded area) compared with simulation predicted values (red) along three routing crystal directions (indicated as black arrows). (d) Measured crosstalk (blue) spectra compared with simulation predicted values (red) along three routing crystal directions.

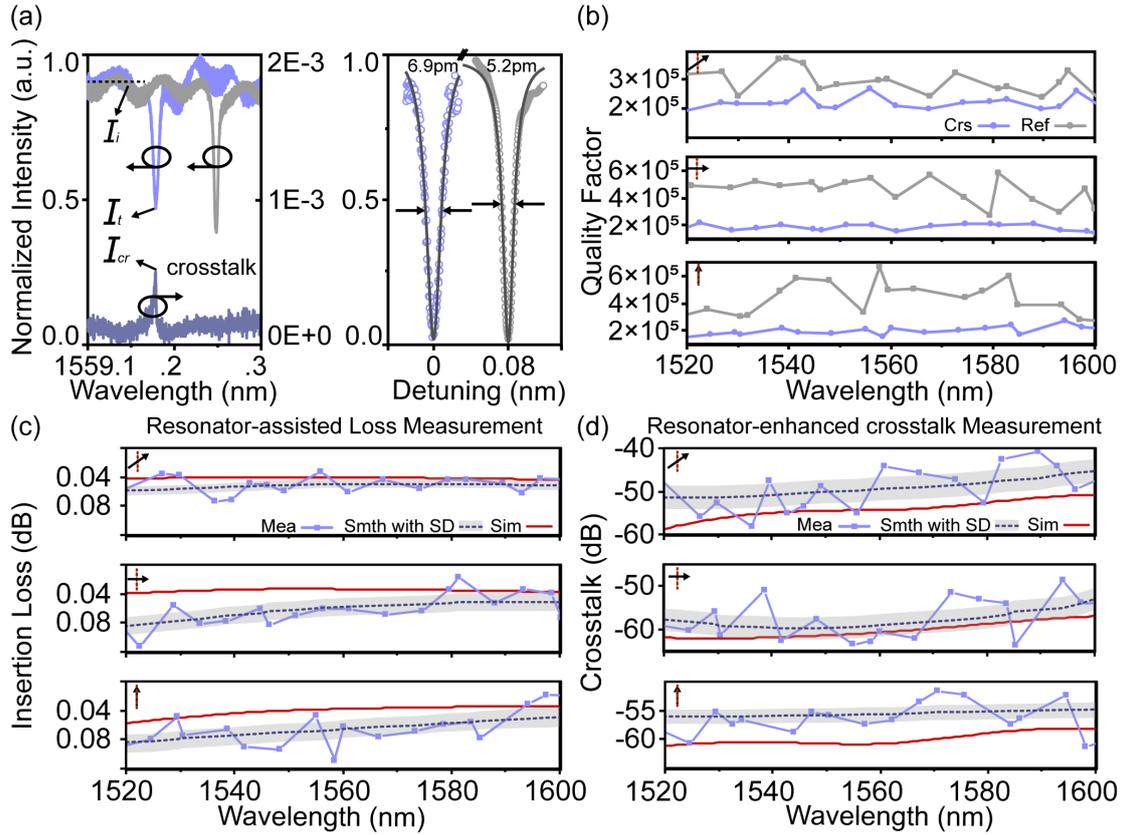

**Fig. 4.** Resonator-assisted characterizations of TFLN crossings. (a) Example normalized transmission spectra of the through port of a crossing-embedded (bright blue) and a reference (gray) resonator, as well as the cross port of the crossing resonator (dark blue). Right: Zoom-in view and Lorentzian-fit of the resonances. (b) Loaded Q factors of reference resonator (gray) and crossing-embedded resonator (blue) extracted at different wavelengths along the three routing crystal directions (black arrows). (c-d) Raw extracted (blue dots), smoothed (dashed blue), and simulated insertion losses (c) and crosstalk (d) at different wavelengths. Shaded areas show standard deviations of the measurements.